\documentclass{ws-ijgmmp}
\usepackage{color}
\begin{document}

\markboth{F. Ahmed and N. Candemir}
{Generalized DKP-oscillator in topological defects}

%
\catchline{}{}{}{}{}
%

\title{Generalized Duffin-Kemmer-Petiau oscillator under Aharonov-Bohm flux in topological defects backgrounds
}

\author{Faizuddin Ahmed\footnote{email: \bf faizuddinahmed15@gmail.com; faizuddin@ustm.ac.in}}

\address{Department of Physics, University of Science \& Technology Meghalaya, Ri-Bhoi, 793101, India
}

\author{Nuray Candemir\footnote{email: \bf ncandemi@eskisehir.edu.tr}}

\address{Department of Physics, Faculty of Science, Eskisehir Technical University, Eskisehir, Turkey}

\maketitle


\begin{abstract}
In this article, we study the generalized Duffin-Kemmer-Petiau (DKP) oscillator under the influence of quantum flux field in the topological defects produced by a cosmic string space-time and point-like global monopole. The generalized DKP oscillator will be investigated through a non-minimal substitution of the momentum operator $\vec{p} \to \left(\vec{p}+i\,M\,\omega\,\eta^0\,f(r)\,\hat{r}\right)$ in the relativistic DKP equation. We solve this generalized DKP oscillator in a cosmic string space-time background and obtain the energy levels and wave function of the oscillator field using the parametric Nikiforov-Uvarov method. Afterwards, we solve the generalized DKP-oscillator in a point-like global monopole space-time and obtain the energy levels and wave functions following the same method. In fact, it is shown there that the energy eigenvalues are influenced by the topological defect of cosmic string and point-like global monopole and gets modified compared to flat space results, and breaks the degeneracy of the energy levels. Furthermore, we observe that the eigenvalue solutions depends on the quantum flux field that shows the gravitational analogue of the Aharonov-Bohm effect and also gives us a persistent currents.
\end{abstract}



\section{Introduction}

In recent times, study of the gravitational field effects on quantum mechanical problems in curved space-times has become a significant area of research, attracting the attention of many researchers. Investigating the interactions between quantum particles (bosons or fermions) and gravity is an active and important field of interest. To understand these interactions, researchers have focused on solving relativistic and non-relativistic wave equations. The study of wave equations in various curved space-time geometries has been extensive. Some notable examples include investigations in G\"{o}del cosmological solutions \cite{KG,ND,YYIJMPA}, Som-Raychaudhuri space-time \cite{ND,SR}, Schwarzschild-like solutions \cite{CC}, and both topologically trivial \cite{FA} and non-trivial \cite{LCNS} space-time backgrounds. Overall, the exploration of wave equations in curved space-times with gravitational interactions and external fields offers valuable insights into the fundamental aspects of quantum mechanics and the interplay between quantum particles and gravity, contributing to the advancement of theoretical physics and our understanding of the universe.

Numerous authors have extensively studied relativistic wave equations in curved space-times with topological defects, as well as in the presence of topological defects produced by cosmic strings and point-like global monopoles. The dynamics of spin-$0$ scalar particles and spin-$\frac{1}{2}$ particles are described by the Klein-Gordon equation and the Dirac equation, respectively. These wave equations have been thoroughly investigated in various backgrounds with topological defects, leading to a wealth of research outcomes. Some notable studies in this area include: Hydrogen atom in cosmic string and point-like global monopole backgrounds \cite{hh1}, spin-$0$ oscillator field under a magnetic field in a cosmic string space-time \cite{AB}, generalized Dirac oscillator under an external magnetic field in a cosmic dislocation space-time \cite{HC}, quantum motions of scalar and spin-half particles under a magnetic field in cosmic string space-time \cite{hh2}, quantum scattering of an electron by a topological defect called dispiration with an externally applied magnetic field \cite{ff1}, Dirac fermions in G\"{o}del-type background space-times with torsion \cite{ff2}, relativistic dynamics of a neutral particle with magnetic dipole moment interacting with an external electric field \cite{ff3}, holonomic quantum computation based on the defect-mediated properties of graphite cones \cite{ff4}, and the relativistic and non-relativistic quantum dynamics of a neutral particle with a permanent magnetic dipole moment interacting with two distinct field configurations in a cosmic string space-time \cite{ff5}. These investigations have provided valuable insights into the behavior and properties of particles in the presence of topological defects, shedding light on their quantum dynamics and interactions in diverse space-time geometries. Additionally, the study of non-relativistic wave equations in the background of topological defects with physical potential models has also been explored in the literature \cite{COE2,fa2}. This research expands the understanding of the behavior of non-relativistic particles in various physical scenarios involving topological defects. Overall, the investigation of wave equations in the presence of topological defects and in curved space-times has proven to be a rich area of research with wide-ranging implications for various branches of physics, including particle physics, quantum mechanics, and cosmology.   

There exists another relativistic wave equation known as the Duffin-Kemmer-Petiau (DKP) equation, which is a first-order wave equation similar to the Dirac equation. The DKP equation describes both spin-$0$ and spin-$1$ fields or particles \cite{bb1,bb2,bb3,bb4}, making it a versatile framework for studying relativistic interactions of such particles. Notably, the DKP equation is a direct generalization of the Dirac equation based on the DKP algebra \cite{WG}. The DKP equation has been the subject of investigation by various authors in different physical scenarios. Some notable studies include: scalar bosons in the presence of the Aharonov-Bohm (AB) flux field and Coulomb potential in cosmic string and global monopole backgrounds \cite{aa1}, quantum dynamics of spin-$0$ particles described by the DKP equation in a G\"{o}del-type space-time \cite{hh3}, DKP equation in a magnetic cosmic string background \cite{hh5}, DKP equation in cosmic string backgrounds under a magnetic field in $1+2$ dimensions \cite{MD}, DKP equation in a cosmic-string space-time with the Cornell interaction \cite{MM}, and the DKP equation in a topologically trivial space-time \cite{hh4}. Additionally, the DKP equation has been utilized in analyzing relativistic interactions of spin-zero and spin-one hadrons with nuclei \cite{NY2}, showcasing its significance in particle physics and nuclear physics research. These investigations demonstrate the versatility and broad applicability of the DKP equation in various physical contexts, shedding light on the behavior and properties of particles with different spins in diverse space-time backgrounds and potential interactions.

Indeed, oscillator of the DKP equation has been extensively studied in various curved space-times and backgrounds involving topological defects, with multiple investigations in the existing literature. The DKP oscillator is characterized by replacing the momentum vector $\vec{p} \to (\vec{p}+i\,M\,\omega\,\eta^0\,\vec{r})$ \cite{NY} in the DKP equation, where $\omega$ represents the oscillator frequency, $\eta^0=\Big(2\,(\beta^0)^2-1\Big)$, and $r$ denotes the radial distance of the particle from the symmetry axis. The DKP oscillator equation has been analyzed in several contexts, including: cosmic string space-time \cite{hh6}, non-inertial effects in cosmic string space-time \cite{hh7}, minimal length effects \cite{hh8,hh9}, noncommutative phase space \cite{hh10,hh11,hh12,hh13}, Dunkl derivative context \cite{hh14}, linear interaction in cosmic string space-time \cite{hh15}, spinning cosmic string space-time \cite{hh16}, presence of Coulomb potential in cosmic string space-time in 2D \cite{hh17,EPJP}, one-dimensional systems \cite{hh18}, cosmic screw dislocation background \cite{SZ}, background space-time around a chiral cosmic string \cite{NC}, Som-Raychaudhuri space-time \cite{YY2}, topologically trivial space-time in 4D \cite{FA2}. Furthermore, the generalized oscillator of the DKP equation has also been thoroughly investigated in the literature, with notable examples including: generalized Kemmer oscillator in a cosmic string background under a magnetic field in $1+2$ dimensions \cite{hh19}, generalized Kemmer oscillator in $1D$ \cite{hh20}, generalized DKP oscillator with linear, Coulomb, and Cornell potential functions in a cosmic string space-time \cite{SZ2}, generalized DKP oscillator for spin-$0$ particles in a spinning cosmic string space-time \cite{YY}, relativistic generalized boson oscillator in a chiral conical space-time background \cite{aa5}. In these generalized cases, the DKP oscillator is analyzed using the non-minimal substitution $\vec{p} \to (\vec{p}+i\,M\,\omega\,\eta^0\,\vec{r}) \to (\vec{p}+i\,M\,\omega\,\eta^0\,f(r)\,\hat{r})$, where $f(r)$ is an arbitrary function. For this particular analysis, we select the Cornell-type potential function (linear plus a Coulomb-type function) represented by $f(r)=\Big(\zeta\,r+\frac{\delta}{r}\Big)$, where $\zeta$ and $\delta$ are positive integer constants. This choice allows us to study the dynamics of scalar bosons via the generalized DKP oscillator in the given context. These comprehensive investigations contribute significantly to our understanding of the behavior and properties of the DKP oscillator and its generalized version in various physical scenarios, providing valuable insights into the quantum dynamics of particles in curved space-times, topological defects, and backgrounds with potential interactions.

The primary motivation for this work stems from the outcomes presented in Ref. \cite{hh6}. In that study, authors explored the DKP oscillator in the background of a cosmic string space-time in the cylindrical coordinates described by the metric $ds^2=-dt^2+dr^2+\alpha^2\,r^2\,d\phi^2+dz^2$. They were able to obtained analytical solutions for the quantum system. Additionally, in Ref. \cite{hh7}, the DKP oscillator within a rotating cosmic string background were investigated and analytically solved the wave equation. In Ref. \cite{aa1}, the DKP equation was studied in the presence of a Coulomb potential and an Aharonov-Bohm flux field in cosmic string and point-like global monopole space-times. As a natural extension of these previous works, we direct our attention to the quantum dynamics of scalar bosons through the generalized DKP oscillator, embedded in the background of a cosmic string space-time, while considering the influence of a quantum flux field. Subsequently, we also investigate the same problem in the context of a point-like global monopole space-time, with the influence of the flux field. This flux field is introduced by means of a minimal substitution in the relativistic wave equation. Importantly, it should be noted that the quantum dynamics of scalar bosons via the DKP oscillator or its generalization in a point-like global monopole background has not yet been addressed in those Refs. \cite{aa1,hh6}, which constitutes the primary focus of our current work. In this paper, we employ the parametric Nikiforov-Uvarov method to solve the generalized DKP oscillator, a technique widely recognized for its efficacy in solving quantum mechanical problems, as demonstrated in various literature references (see, Refs. \cite{SZ2,hh20,aa7}). Through our approach, we not only derive solutions for the wave equation but also analyze the effects of the topological defect and quantum flux field on the resulting eigenvalue solutions. Our findings differ from those obtained in Refs. \cite{hh6,hh7} concerning the cosmic string space-time in the cylindrical system and Ref. \cite{aa1}. Overall, this work presents a comprehensive exploration of the quantum dynamics of scalar bosons via the generalized DKP oscillator in a point-like global monopole background, which fills a significant gap in the existing literature.

This paper is structured as follows: Section 1 provides an introduction to the topic and outlines the objectives of the study. In Section 2, we delve into the DKP equation, particularly focusing on its behavior in a curved space background. Moving on to Section 3, we explore the generalized DKP oscillator in the context of a cosmic string space-time, taking into account the presence of a quantum flux field. In Section 4, we extend our investigation to the generalized DKP oscillator in a point-like global monopole scenario under the influence of a quantum flux field. Section 5 is dedicated to the examination of the persistent currents in the quantum systems under consideration. Finally, in Section 6, we draw our conclusions based on the findings from the previous sections. Throughout our analysis, we adopt a system of units where $c=1=\hbar=G$, simplifying the equations and calculations.

\section{The Duffin-Kemmer-Petiau equation in curved geometry background}

In this section, we will study the DKP equation in curved space-time in the presence of the electromagnetic four-vector potential $A_{\mu}$. We consider the topological defect of a cosmic string space-time and point-like global monopole and derived the necessary quantities involved in the wave equation.

The first-order relativistic DKP equation for a charged free scalar bosons of mass $M$ in flat space is given by \cite{bb1,bb2,bb3,bb4}
\begin{equation}
\Big(i\,\beta^{\mu}\,\partial_{\mu}-M\Big)\,\Psi=0,
\label{b1}
\end{equation}
where $\beta^{\mu}$ are the DKP matrices which satisfy the following commutation rules:
\begin{equation}
\beta^{\mu}\,\beta^{\nu}\,\beta^{\sigma}+\beta^{\sigma}\,\beta^{\nu}\,\beta^{\mu}=\eta^{\mu\nu}\,\beta^{\sigma}+\eta^{\sigma\nu}\,\beta^{\mu}.
\label{b2}
\end{equation}
Here $\eta^{\mu\nu}=\mbox{diag} (-1,+1,+1,+1)$ is the Minkowski metric tensor. 

In curved space-time with an electromagnetic four-vector potential $A_{\mu}$, this DKP equation can be written as 
\begin{equation}
\Big[i\,\beta^{\mu}({\bf x})\,(\nabla_{\mu}-i\,q\,A_{\mu})-M\Big]\,\Psi=0,
\label{b3}
\end{equation}
where $\beta^{\mu} ({\bf x})=e^{\mu}_{\,a} ({\bf x})\,\beta^a$ is the DKP matrices in curved space-time that satisfy the relation (\ref{b2}) by replacing $\eta_{\mu\nu} \to g_{\mu\nu}$, $q$ is the electric charges and $e^{\mu}_{a} ({\bf x})$ is the tetrad basis vector obeying the following relations
\begin{eqnarray}
&&e^{\mu}_{\,a} ({\bf x})\,e^{a}_{\nu} ({\bf x})=\delta^{\mu}_{\nu},\quad e^{a}_{\mu} ({\bf x})\,e^{\mu}_{\,b} ({\bf x})=\delta^{a}_{b},\quad g_{\mu\nu} ({\bf x})=e^{a}_{\mu}({\bf x})\,e^{b}_{\nu} ({\bf x})\,\eta_{ab},\nonumber\\
&&\eta_{ab}=e^{\mu}_{\,a} ({\bf x})\,e^{\nu}_{\,b} ({\bf x})\,g_{\mu\nu} ({\bf x}).
\label{b4}
\end{eqnarray}
And the covariant derivative $\nabla_{\mu}$ is defined by \cite{JTL,SW}
\begin{equation}
\nabla_{\mu}=\partial_{\mu}+\Gamma_{\mu}\quad ,\quad \Gamma_{\mu}=\frac{1}{2}\,\omega_{\mu\,ab}\,[\beta^a, \beta^b].
\label{b5}
\end{equation}
Here $\omega_{\mu\,ab}$ is the spin-connection defined by
\begin{eqnarray}
\omega^{\,\,\,\,a}_{\mu\,\,\,\,b}=e^{a}_{\tau}\,e^{\nu}_{\,\,b}\,\Gamma^{\tau}_{\mu\nu}-e^{\nu}_{\,\,b}\,\partial_{\mu}\,e^{a}_{\nu},\quad
\Gamma^{\tau}_{\mu\nu}=\frac{1}{2}\,g^{\tau\sigma}\,(g_{\nu\sigma,\mu}+g_{\mu\sigma,\nu}-g_{\mu\nu,\sigma}).
\label{b6}
\end{eqnarray}

Below, we will calculate the spin connection in a cosmic string space-time and point-like global monopole background.

\subsection{\bf Cosmic string space-time}

Here, we will consider the cosmic string space-time and determine all the above quantities, and finally derive the wave equation. The cosmic string space-time in the spherical system is given by \cite{aa1,cc12,cc14,cc8,aa2,ff6}
\begin{equation}
ds^2=-dt^2+dr^2+r^2\,(d\theta^2+a^2\,\sin^2 \theta\,d\phi^2),
\label{b7}
\end{equation}
where $a$ is the topological defect parameter. The metric tensors for cosmic string space-time are 
\begin{eqnarray}
    g_{\mu\nu}=\begin{pmatrix}
        -1~~~~~&0~~~&0~~~&0\\
        0~~~&1~~~&0~~~&0\\
        0~~~&0~~~&r^2~~&0\\
        0~~~&0~~~&0~~~&a^2\,r^2\,\sin^2 \theta
    \end{pmatrix},\quad
    g^{\mu\nu}=\begin{pmatrix}
        -1~~~~~&0~~~&0~~~&0\\
        0~~~&1~~~&0~~~&0\\
        0~~~&0~~~&\frac{1}{r^2}~~&0\\
        0~~~&0~~~&0~~~&\frac{1}{a^2\,r^2\,\sin^2 \theta}
    \end{pmatrix}.
    \label{metric}
\end{eqnarray}

For the cosmic string geometry, the tetrad basis vector $e^{\mu}_{a}$ is defined by
\begin{eqnarray}
e^{\mu}_{\,a}({\bf x})=
\begin{pmatrix}
1~~~ &0~~~&0~~~&0\\
0~~~ &1~~~&0~~~&0\\
0~~~ &0~~~&\frac{1}{r}~~& 0\\
0~~~ &0~~~&0~~~&\frac{1}{a\,r\,\sin \theta}
\end{pmatrix},\quad
e^{a}_{\mu}({\bf x})=
\begin{pmatrix}
1~~~ &0~~~&0~~~&0\\
0~~~ &1~~~&0~~~&0\\
0~~~ &0~~~&r~~~&0\\
0~~~ &0~~~&0~~~& a\,r\,\sin \theta
\end{pmatrix}.
\label{b8}
\end{eqnarray}

The Christoffel symbols are
\begin{eqnarray}
&&\Gamma^{r}_{\mu\nu}=-r
\begin{pmatrix}
0~&0~&0~&0\\
0~&0~&0~&0\\
0~&0~&1~&0\\
0~&0~&0~&a^2\,\sin^2 \theta\\
\end{pmatrix},\quad
\Gamma^{\theta}_{\mu\nu}=
\begin{pmatrix}
0~~&0~~&0~~&0\\
0~~&0~~&\frac{1}{r}~~&0\\
0~~&\frac{1}{r}~~&0~~~&0\\
0~~&0~~&0~~&-\frac{a^2\,\sin 2\theta}{2}
\end{pmatrix},\nonumber\\
&&\Gamma^{\phi}_{\mu\nu}=
\begin{pmatrix}
0~~&0~~&0~~&0\\
0~~&0~~&0~~&\frac{1}{r}\\
0~~&0~~&0~~~&\cot \theta\\
0~~&\frac{1}{r}~~&\cot \theta~~&0
\end{pmatrix}.
\label{b10}
\end{eqnarray}

The spin-connection of the cosmic string space-time using the tetrad basis (\ref{b8}) and the Christoffel symbols $\Gamma^{\sigma}_{\mu\nu}$ given in (\ref{b10}) are as follows:
\begin{eqnarray}
\omega_{\theta\,\,ab}=
\begin{pmatrix}
0~~~~&0~~~ &0~~~ & 0~~~\\
0~~~~&0~~~ &1~~~ & 0~~~\\
0~~~~ &-1~~~~~&0~~~ & 0~~~\\
0~~~~&0~~~&0~~~ & 0~~~
\end{pmatrix},\quad
\omega_{\phi\,\,ab}=
\begin{pmatrix}
0~~~ &0~~~&0~~~&0\\
0~~~ &0~~~&0~~~&a \sin \theta\\
0~~~ &0~~~&0~~~&a\cos \theta \\
0~~~ &-a\sin \theta~~~ &-a\cos \theta~~~ &0
\end{pmatrix}.
\label{b9}
\end{eqnarray}

\subsection{\bf Point-like global monopole space-time}

Here, we consider the point-like global monopole space-time and determine all the necessary quantities and will derive the wave equation. This space-time geometry in the spherical system is given by \cite{aa1, aa2, ff6, cc11, ff7, ff8, ff9, ff10, ff11}
\begin{equation}
ds^2=-dt^2+dr^2+b^2\,r^2\,(d\theta^2+\sin^2 \theta\,d\phi^2),
\label{c1}
\end{equation}
where $b$ is the topological defect parameter. The properties of this space-time geometry were discussed in details in Refs. \cite{cc1,cc3,cc5}.  

The metric tensors for the point-like global monopole space-time are
\begin{eqnarray}
    g_{\mu\nu}=\begin{pmatrix}
        -1~~~~~&0~~~&0~~~&0\\
        0~~~&1~~~&0~~~&0\\
        0~~~&0~~~&b^2\,r^2~~&0\\
        0~~~&0~~~&0~~~&b^2\,r^2\,\sin^2 \theta
    \end{pmatrix},\quad
    g^{\mu\nu}=\begin{pmatrix}
        -1~~~~~&0~~~&0~~~&0\\
        0~~~&1~~~&0~~~&0\\
        0~~~&0~~~&\frac{1}{b^2\,r^2}~~&0\\
        0~~~&0~~~&0~~~&\frac{1}{b^2\,r^2\,\sin^2 \theta}
    \end{pmatrix}.
    \label{metric2}
\end{eqnarray}

The Christoffel symbols of this space-time are given by
\begin{eqnarray}
&&\Gamma^{r}_{\mu\nu}=-r\,b^2
\begin{pmatrix}
0~&0~&0~&0\\
0~&0~&0~&0\\
0~&0~&1~&0\\
0~&0~&0~&\,\sin^2 \theta\\
\end{pmatrix},\quad
\Gamma^{\theta}_{\mu\nu}=
\begin{pmatrix}
0~~&0~~&0~~&0\\
0~~&0~~&\frac{1}{r}~~&0\\
0~~&\frac{1}{r}~~&0~~~&0\\
0~~&0~~&0~~&-\frac{\sin 2\theta}{2}
\end{pmatrix},\nonumber\\
&&\Gamma^{\phi}_{\mu\nu}=
\begin{pmatrix}
0~~&0~~&0~~&0\\
0~~&0~~&0~~&\frac{1}{r}\\
0~~&0~~&0~~~&\cot \theta\\
0~~&\frac{1}{r}~~&\cot \theta~~&0
\end{pmatrix}.
\label{c4}
\end{eqnarray}

For the point-like global monopole, the non-zero components of the Riemann tensor $R^{\lambda}_{\mu\nu\sigma}$ are $R^{\theta}_{\phi\phi\theta}=(-1+b^2)\,\sin^2 \theta$ and $R^{\phi}_{\theta\phi\theta}=(1-b^2)$. Also, the Ricci scalar is given by $R=R^{\mu}_{\,\mu}=\frac{2\,(1-b^2)}{b^2\,r^2}$. One can see that for $b \to 1$, the space-time reduces to Minkowski flat space in the spherical system. This geometry possesses a conical singularity at $r=0$.

For the point-like global monopol, the tetrad basis vector $e^{\mu}_{a}$ is defined by
\begin{eqnarray}
e^{\mu}_{\,a}({\bf x})=
\begin{pmatrix}
1~~~ &0~~~ &0~~~ & 0\\
0~~~ &1~~~ &0~~~ & 0\\
0~~~ &0~~~ &\frac{1}{b\,r}~~~&0\\
0~~~ &0~~~ &0~~~ &\frac{1}{b\,r\,\sin \theta}
\end{pmatrix},\quad
e^{a}_{\mu}({\bf x})=
\begin{pmatrix}
1~~~ &0~~~ &0~~~&0\\
0~~~ &1~~~ &0~~~&0\\
0~~~ &0~~~ &b\,r~~&0\\
0~~~ &0~~~ &0~~~ &b\,r\,\sin \theta
\end{pmatrix}.
\label{c2}
\end{eqnarray}

The spin-connection of the cosmic string space-time using the tetrad basis vector (\ref{b8}) and the Christoffel symbols $\Gamma^{\sigma}_{\mu\nu}$ given in (\ref{c4}) are
\begin{eqnarray}
\omega_{\theta\,\,ab}=
\begin{pmatrix}
0~~~~&0~~~ &0~~~ & 0~~~\\
0~~~~&0~~~ &b~~~ & 0~~~\\
0~~~~ &-b~~~~~&0~~~ & 0~~~\\
0~~~~&0~~~&0~~~ & 0~~~
\end{pmatrix},\quad
\omega_{\phi\,\,ab}=
\begin{pmatrix}
0 & 0 & 0 & 0\\
0 & 0 & 0 & b \sin \theta\\
0 & 0 & 0 & \cos \theta \\
0 & -b \sin \theta & -\cos \theta &0
\end{pmatrix}.
\label{c3}
\end{eqnarray}

\section{Effects of AB-flux on generalized DKP-oscillator in cosmic string space-time}

In this section, we study the relativistic quantum motions of scalar massive charged bosons in the topological defects background under the influence of the Aharonov-Bohm flux field. The DKP oscillator is carrying out by a non-minimal substitution in the DKP equation via \cite{NY,hh6}
\begin{equation}
\label{eq.3}
p_{\mu}\rightarrow p_{\mu}+i\,M\,\omega\,\eta^{0}\,X_{\mu},
\end{equation}
where $p_{\mu}=-i\,\nabla_{\mu}$ is the four-momentum operator with covariant derivative $\nabla_{\mu}\equiv\partial_{\mu}+\Gamma_{\mu}$. In Eq. (3), $\omega$ is the angular frequency of the DKP oscillator. To generalize the DKP oscillator, the four-vector $X_{\mu}=(0,r,0,0)$ is replaced by $X_{\mu}=(0,f(r),0,0)$, where $f(r)$ is an arbitrary function. Thus, the generalized DKP oscillator in a curved space-time is written as
\begin{equation}
\label{eq.4}
\Bigg[i\,\beta^{\mu}\,\Big(\partial_{\mu}+\Gamma_{\mu}+M\,\omega\,\eta^{0}\,X_{\mu}-i\,q\,A_{\mu}\Big)-M\Bigg]\,\Psi=0.
\end{equation}

Throughout the analysis, we choose the electromagnetic four-vector potential $A_{\mu}=(0, A_{r}, A_{\theta}, A_{\varphi})$ with the following components
\begin{equation}
\label{eq.5}
A_{r}=0=A_{\theta},\quad A_{\varphi}=\frac{\Phi_{AB}}{2\,\pi\,a\,r\,\sin \theta},
\end{equation}
where $\Phi_{AB}=const$ is the Aharonov-Bohm flux field.

Since the interaction is time-independent, one can write the total wave function as
\begin{equation}
\label{eq.6}
\Psi(t, r, \theta\, \phi)=\exp(-i\,E\,t)\,\psi(r,\theta,\phi),
\end{equation}
where $E$ is the particles energy.

Substituting Eqs. (\ref{eq.5}) and Eq. (\ref{eq.6}) into the Eq.(\ref{eq.4}), one reads
\begin{eqnarray}
\label{eq.7}
&&\Bigg[\beta^{0}\,E+i \beta^{1}(\partial_{r}+M\omega \eta^{0} f(r) )+i\frac{\beta^{2}}{r}(\partial_{\theta}-\beta^{2}\beta^{1})+i\frac{\beta^{3}}{a\,r\,\sin \theta }\Big\{
(\partial_{\varphi}-i\Phi)-a\,\sin \theta \beta^{3}\beta^{1}\nonumber\\
&&-a\,\cos \theta \beta^{3}\beta^{2}\Big\}\Bigg] \psi=M\,\psi,
\end{eqnarray}
where $\Phi=\frac{\Phi_{AB}}{\Phi_0}$, and $\Phi_0=2\,\pi\,q^{-1}$. 

The five-component spinor can be written as
\begin{equation}
\label{eq.8}
\psi^{T}(r,\theta,\phi)=[\psi_{1},....\psi_{5}]
\end{equation}
Then, inserting the above spinor into the Eq.(\ref{eq.7}), the following system of equations are obtained
 \begin{eqnarray}
\label{eq.9}
&&E\,\psi_{2}-i\left(\partial_{r}+\frac{2}{r}-M\omega f(r)\right)\psi_{3}-\frac{i}{r}\left(\partial_{\theta}+\frac{\cos \theta}{\sin \theta}\right)\psi_{4}-\frac{i}{a\,r \sin \theta }\,(\partial_{\varphi}-i\,\Phi)\,\psi_{5}\nonumber\\
&&=M\,\psi_{1}.\\
&&E\,\psi_{1}=M\,\psi_{2}.\\
&&i\,(\partial_{r}+M\omega f(r))\,\psi_{1}=M\,\psi_{3}.\\
&&i\,\partial_{\theta}\,\psi_{1}=M\,\psi_{4}.\\
&&\frac{i}{a\,r\,\sin \theta }\,(\,\partial_{\varphi}-i\,\Phi)\,\psi_{1}=M\,\psi_{4}.
\end{eqnarray}

The above equation system for $\psi_{1}$ gives the following wave equation
\begin{eqnarray}
\label{eq.14}
\Bigg[\frac{d^{2}}{dr^{2}}+\frac{2}{r}\frac{d}{dr}+E^2-M^2\,\omega^2\,f^{2}(r)+\frac{2M \omega f(r)}{r}+M\omega f'(r)-\frac{L^{2}_{eff}}{r^{2}}-M^2\Bigg]\psi_{1}=0,
\end{eqnarray}
where different operators are given by
\begin{eqnarray}
\label{eq.15}
&&L^{2}_{eff}=-\Bigg[\frac{1}{\sin\theta}\,\frac{d}{d\theta}\,\left(\,\sin\theta\frac{d}{d\theta}\,\right)+\frac{1}{a^2\,\sin^{2}\theta }\left(\frac{d}{d\varphi}-i\Phi\right)^2\Bigg],\nonumber\\
&&L^{eff}_{z}=-\frac{i}{a}\,\left(\frac{d}{d\varphi}-i\Phi\right).
\end{eqnarray}
Their eigenvalue equations are as follows:
\begin{eqnarray}
\label{eq.16}
L^{eff}_{z}\,\psi_1=m'\,\psi_1,\quad L^{2}_{eff}\,\psi_1=\lambda'\,\psi_1,\quad m'=\frac{m-\Phi}{a},\quad \lambda'=\ell'\,(\ell'+1),
\end{eqnarray}
where $\ell'\geq |m'|$ that implies $\ell'=(|m'|+\kappa)$ with $\kappa=0,1,2,..$. These eigenvalues depend on the magnetic quantum flux $\Phi$ and the geometric parameter of space $a$. For $a \to 1$, the space-time under consideration becomes Minkowski flat space.

In order to solve Eq.(\ref{eq.14}), it is reasonable to use the following ansatz
\begin{equation}
\label{eq.17}
\psi_{1} (r,\theta,\phi)=R(r)\chi(\theta,\varphi).
\end{equation}

Substituting (\ref{eq.17}) into the Eq. (\ref{eq.14}), one will have
\begin{eqnarray}
\label{eq.18}
\Bigg[\frac{d^{2}}{dr^{2}}+\frac{2}{r}\frac{d}{dr}-\frac{\ell'\,(\ell'+1)}{r^{2}}-M^{2}\omega^{2}f^{2}(r)+M\omega f'(r)+\frac{2 M \omega f(r)}{r}+E^2-M^2 \Bigg]R(r)=0.
\end{eqnarray}

In the literature, several authors have been studied the generalized oscillator field by solving the Klein-Gordon equation, the Dirac equation and the DKP equation in different space-time backgrounds. For the studies of the generalized DKP oscillator in the considered system, we chose the function $f(r)$ the linear plus Coulomb-type given by :
\begin{equation}
\label{eq.19}
f(r)=\Big(\zeta\,r+\frac{\delta}{r}\Big),
\end{equation}
where $\zeta, \delta$ are positive constants. For $\delta \to 0$, the function becomes linear in $r$ and the quantum system is called the DKP oscillator. For $\zeta \to 0$, the function becomes Coulomb-type $f (r) \sim \frac{1}{r}$. This type of function $f(r)$ given in Eq. (\ref{eq.19}) in the literature known as the Cornell-type potential form function that has widely been used by many authors \cite{aa1,aa2,aa5,SZ2,aa10,aa11,aa12}.

Substituting this function (\ref{eq.19}) in the Eq (\ref{eq.18}), we have
\begin{equation}
\label{eq.20}
\Bigg[\frac{d^{2}}{dr^{2}}+\frac{2}{r}\frac{d}{dr}-M^2\,\omega^2\,\zeta^2\,r^2-\frac{j^2-\frac{1}{4}}{r^{2}}+\Lambda\Bigg]\,R(r)=0.
\end{equation}
Defining a new function by $R(r)=r^{-1/2}\,U(r)$ and using a new variable $x=M\,\omega\,\zeta\,r^2$, one will obtain the following differential equation
\begin{equation}
\label{eq.21}
U''(x)+\frac{1}{x}\,U'(x)+\frac{1}{x^2}\,\left(-\xi_1\,x^2+\xi_2\,x-\xi_3\right)\,U(x)=0,
\end{equation}
where
\begin{eqnarray}
\label{eq.22}
&&\xi_1=\frac{1}{4}, \quad \xi_2=\frac{\Lambda}{4\,M\,\omega\,\zeta},\quad \xi_3=\frac{j^2}{4},\quad \Lambda=E^2-M^2+3\,M\,\omega\,\zeta-2\,M^2\,\omega^2\,\zeta\,\delta,\nonumber\\
&&j=\sqrt{\ell'\,(\ell'+1)+\Big(M\,\omega\,\delta-\frac{1}{2}\Big)^2}.
\end{eqnarray}

Equation (\ref{eq.21}) is a second-order homogeneous differential equation that can be solved using the well-known parametric Nikiforov-Uvarov method \cite{AFN}. This NU method has renowned interest in solving the quantum mechanical problems in the literature. Several authors have been solved the relativistic and non-relativistic wave equations using this method and obtained the eigenvalue solutions (see, Refs. \cite{SZ2,hh20}). 

Thereby, comparing (\ref{eq.21}) with Eq. (A.1) in appendix in Ref. \cite{aa7}, we have $c_1=1$, $c_2=0=c_3$ and other coefficients are
\begin{eqnarray}
&&c_4=0,\quad c_5=0,\quad c_6=\xi_1,\quad c_7=-\xi_2,\quad c_8=\xi_3,\quad c_9=\xi_1,\quad c_{10}=1+2\,\sqrt{\xi_3},\nonumber\\
&&c_{11}=2\,\sqrt{\xi_1},\quad c_{12}=\sqrt{\xi_3},\quad c_{13}=-\sqrt{\xi_1}.
\label{eq.23}
\end{eqnarray}

Substituting Eq. (\ref{eq.23}) in Eq. (A.3) in appendix in Ref. \cite{aa7}, we obtain the following energy eigenvalue expression
\begin{equation}
E_{n,m}=\pm\,\sqrt{M^2+M\,\omega\,\zeta\,\Bigg(M\,\omega\,\delta+4\,n+2\,j-1\Bigg)},
\label{eq.24}
\end{equation}
where radial quantum number is defined by $n=0,1,2,3,...$ and $j$ is given in (\ref{eq.22}) by
\begin{equation}\nonumber
j=\sqrt{\Big(\frac{|m-\Phi|}{a}+\kappa\Big)\Big(\frac{|m-\Phi|}{a}+\kappa+1\Big)+\Big(M\,\omega\,\delta-\frac{1}{2}\Big)^2}.
\end{equation}

The energy eigenvalue (\ref{eq.24}) shows that the discrete set of the generalized DKP oscillator energies are symmetrical about $E_{n,m}=0$ and this is irrespective of the sign of $m$. This fact is associated to the fact that the generalized DKP oscillator embedded in a cosmic string space-time background does not distinguish particles from antiparticles.

The radial wave function will be
\begin{equation}
U_{n,m} (x)=N_{n,m}\,x^{\frac{j}{2}}\,e^{-\frac{x}{2}}\,L^{(j)}_{n} (x),
\label{eq.25}
\end{equation}
where $N_{n,m}$ is the normalization constant.

In terms of $r$, the radial wave function will be
\begin{equation}
R_{n,m} (r)=N_{n,m}(M\omega\zeta)^{j/2}r^{j-\frac{1}{2}}e^{-\frac{1}{2}M\omega\zeta r^2}L^{(j)}_{n} (M\omega\zeta r^2).
\label{eq.26}
\end{equation}

Equation (\ref{eq.24}) is the relativistic energy spectrum and Eq. (\ref{eq.26}) is the radial wave function of scalar charged bosons via the generalized DKP oscillator equation embedded in the background of a cosmic string space-time under the influence of the AB-flux field. One can see that the eigenvalue solutions are influenced by the topological defect parameter $a$ and the Cornell-type potential function Eq. (\ref{eq.19}). This eigenvalue solution is different from the results of the generalized Dirac oscillator obtained in the literature \cite{aa9} and also different from those results obtained in \cite{aa1,hh6}.

In the function $f(r)$, we choose $\zeta \rightarrow 1$ and $\delta \rightarrow 0$, that is, the DKP oscillator in the cosmic string space-time with the AB-flux field, then the energy eigenvalues from Eq. (\ref{eq.24}) in this special case become
\begin{equation}
\label{eq.30}
E_{n,m}=\pm\,\sqrt{M^2+2\,M\,\omega\,\Bigg(2n+\frac{|m-\Phi|}{a}+\kappa\Bigg)}.
\end{equation}
The normalized radial wave function will be 
\begin{eqnarray}
R_{n,m} (x)=N_{n,m}\,(M\,\omega)^{3/4} x^{\frac{l'}{2}}\, e^{-\frac{x}{2}}\,L^{(l'+1/2)}_{n} (x).
\label{eq.31}
\end{eqnarray}
where $l'$ is given in Eq. (\ref{eq.16}).

This eigenvalue solution Eqs. (\ref{eq.30})--(\ref{eq.31}) reduces to those result obtained in Ref. \cite{hh6} provided zero magnetic flux field $\Phi_{AB} \to 0$ here. Thus, one can see that this eigenvalue solution is shifted by the flux field and gets modified compared to the result in Ref. \cite{hh6}. Hence, we conclude that the relativistic energy spectrum presented in this section by Eq. (\ref{eq.24}) and radial wave function by Eq. (\ref{eq.26}) of the DKP-oscillator are the modified or generalized result to those of obtained in Ref. \cite{hh6} and goes beyond.

\section{Effects of AB-flux on generalized DKP oscillator in global monopole}

In this section, we study the relativistic quantum motions of scalar charged bosons via the generalized DKP oscillator mentioned earlier in a point-like global monopole background under the influence of the AB-flux field. Here, the four-vector of electromagnetic potential given by $A_{\mu}=(0,0,0,A_{\varphi})$ takes the following form
\begin{equation}
\label{eq.42}
A_{\varphi}=\frac{\Phi_{AB}}{2\,\pi\,b\,r\,\sin \theta}.
\end{equation}

Since, the wave function given in Eq. (\ref{eq.6}) for the time-independent interaction is taken as $\Psi(t,r, \theta, \phi)=exp\,(-i\,E\,t)\,\Xi(r, \theta, \phi)$, the generalized DKP oscillator in the presence of the Aharonov-Bohm flux field (\ref{eq.42}) in point-like global monopole defect can be written as
\begin{eqnarray}
\label{eq.43}
&&\Bigg[\beta^{0}\,E+i\,\beta^{1}(\partial_{r}+M\omega \eta^{0} f(r) )+\frac{i\beta^{2}}{b^2\, r}(\partial_{\theta}-\beta^{2}\beta^{1})+\frac{i\beta^{3}}{b^2\,r \,\sin \theta}\Big\{(\partial_{\varphi}-i\,\Phi\,)-b^2\,\sin \theta \beta^{3}\beta^{1}\nonumber\\
&&-\cos \theta \beta^{3}\beta^{2} \Big\}\Bigg]\Xi=M \Xi.
\end{eqnarray}

Substituting the five-component spinor $\Xi^{T} (r, \theta, \phi)=[\Xi_{1},....\Xi_{5}]$ into Eq. (\ref{eq.43}), the generalized DKP oscillator leads to the following five equations:
 \begin{eqnarray}
\label{eq.44}
&&E\,\Xi_{2}-i\left(\partial_{r}+\frac{2}{r}-M\omega f(r)\right)\Xi_{3}-\frac{i}{b^2\,r}\left(\partial_{\theta}+\frac{\cos \theta}{\sin \theta}\right)\Xi_{4}+\frac{i}{b^2\,r\,\sin \theta }\,(\,\partial_{\varphi}-i\,\Phi)\,\Xi_{5}\nonumber\\
&&=M\,\Xi_{1}.\\
\label{eq.45}
&&E\,\Xi_{1}=M\,\Xi_{2}.\\
\label{eq.46}
&& i\,(\partial_{r}+M\omega f(r)\,)\,\Xi_{1}=M\,\Xi_{3}.\\
\label{eq.47}
&& \frac{i}{b^2\, r}\,\partial_{\theta}\,\Xi_{1}=M\,\Xi_{4}.\\
\label{eq.48}
&& \frac{i}{b^2\, r\,\sin \theta }\,(\,\partial_{\varphi}-i\,\Phi) \,\Xi_{1}=M\,\Xi_{4}.
\end{eqnarray}

Combining these equations, one will have the following equation
\begin{eqnarray}
\label{eq.49}
\Bigg[\frac{d^{2}}{dr^{2}}+\frac{2}{r}\frac{d}{dr}+E^2-M^2 \omega^{2}f^{2}(r)+\frac{2M \omega f(r)}{r}+M\omega f'(r)-\frac{L^2}{b^2\,r^2}-M^2\Bigg]\Xi_{1}=0,
\end{eqnarray}
where
\begin{eqnarray}
\label{eq.50}
L^{2}=-\Bigg[\frac{1}{\sin\theta}\,\frac{d}{d\theta}\,\left(\,\sin\theta\,\frac{d}{d\theta}\,\right)+\frac{1}{\sin^{2}\theta }\left(\frac{d}{d\varphi}-i\Phi\right)^2\Bigg], \quad L_{z}=-i\left(\frac{d}{d\varphi}-i\Phi\right).
\end{eqnarray}
The eigenvalue equations of these are
\begin{eqnarray}
\label{eq.51}
&&L^{eff}_{z}\,\Xi_1=m_0\,\Xi_1,\quad m_0=(m-\Phi),\quad L^{2}_{eff}\,\Xi_1=\lambda_0\,\Xi_1, \quad \lambda_0=\ell_0\,(\ell_0+1),\nonumber\\
&&\ell_0=|m_0|+\kappa,
\end{eqnarray}
where $\Phi=\frac{\Phi_{AB}}{\Phi_0}$, $\Phi_0=2\,\pi\,q^{-1}$, $\kappa=0,1,2,...$, $m_0$ and $\lambda_0$ are the eigenvalues of $L_{z}$ and $L^2$ respectively that depends on the magnetic flux $\Phi$.

In order to solve Eq.(\ref{eq.49}), the following ansatz can be used
\begin{equation}
\label{eq.52}
\quad\quad\quad \Xi_{1}=R(r)\,\chi(\theta,\varphi).
\end{equation}
\quad\quad Thus, Eq. (\ref{eq.49}) transforms into
\begin{eqnarray}
\label{eq.53}
\Bigg[\frac{d^{2}}{dr^{2}}+\frac{2}{r}\frac{d}{dr}-\frac{\ell_0\,(\ell_0+1)}{b^2\,r^2}-M^{2}\omega^{2}f^{2}(r)+M\omega f'(r)+\frac{2\,M\,\omega\,f(r)}{r}+E^2-M^2\Bigg]R(r)=0.
\end{eqnarray}

After substituting the function $f(r)$  given in Eq.(\ref{eq.19}), we obtain
\begin{equation}
\label{eq.bb1}
\Bigg[\frac{d^{2}}{dr^{2}}+\frac{2}{r}\frac{d}{dr}-M^2\,\omega^2\,\zeta^2\,r^2-\frac{\tau^2-\frac{1}{4}}{r^{2}}+\Lambda\Bigg]\,R(r)=0.
\end{equation}
Defining a new function 
\begin{equation}
\label{eq.54}
R(r)=r^{-1/2}\,U
\end{equation}
and using a new variable $x=M\,\omega\,\zeta\,r^2$, the formula results in the following differential equation
 \begin{equation}
\label{eq.55}
U''(x)+\frac{1}{x}\,U'(x)+\frac{1}{x^2}\,\Big(-\eta_1\,x^2+\eta_2\,x-\eta_3\Big)\,U(x)=0,
\end{equation}
where
\begin{eqnarray}
\label{eq.56}
&&\eta_1=\frac{1}{4}\quad,\quad \eta_2=\frac{\Lambda}{4\,M\,\omega\,\zeta}\quad,\quad \eta_3=\frac{\tau^2}{4},\nonumber\\ &&\tau=\sqrt{\frac{\ell_0\,(\ell_0+1)}{b^2}+\Bigg(M\,\omega\,\delta-\frac{1}{2}\Bigg)^2},\quad \Lambda=E^2-M^2+3\,M\,\omega\,\zeta-2\,M^2\,\omega^2\,\zeta\,\delta.
\end{eqnarray}

Equation (\ref{eq.55}) can be solved using the same method as stated earlier. Thereby, comparing (\ref{eq.55}) with Eq. (A.1) in appendix in Ref. \cite{aa7}, we have $c_1=1$, $c_2=0=c_3$ and
\begin{eqnarray}
&&c_4=0,\quad c_5=0,\quad c_6=\eta_1,\quad c_7=-\eta_2,\quad c_8=\eta_3,\quad c_9=\eta_1,\quad c_{10}=1+2\,\sqrt{\eta_3},\nonumber\\
&&c_{11}=2\,\sqrt{\eta_1},\quad c_{12}=\sqrt{\eta_3},\quad c_{13}=-\sqrt{\eta_1}.
\label{eq.57}
\end{eqnarray}

Substituting Eq (\ref{eq.57}) in Eq. (A.3) in appendix in Ref. \cite{aa7}, we obtain the following eigenvalue expression
\begin{equation}
E_{n,m}=\pm\,\sqrt{M^2+M\,\omega\,\zeta\,(M\,\omega\,\delta+4\,n+2\,\tau-1)},
\label{eq.58}
\end{equation}
where $\tau$ is given in (\ref{eq.56}) by
\begin{equation}\nonumber
\tau=\sqrt{\frac{(|m-\Phi|+\kappa)\,(|m-\Phi|+\kappa+1)}{b^2}+\Bigg(M\,\omega\,\delta-\frac{1}{2}\Bigg)^2}.
\end{equation}

The energy eigenvalue (\ref{eq.58}) shows that the discrete set of generalized DKP oscillator energies are symmetrical about $E_{n,m}=0$ and this is irrespective of the sign of $m$. This fact is associated to the fact that the generalized DKP oscillator embedded in a point-like global monopole background does not distinguish particles from antiparticles.

The radial wave function will be
\begin{equation}
U_{n,m} (x)=N_{n,m}\,x^{\frac{\tau}{2}}\,e^{-\frac{x}{2}}\,L^{(\tau)}_{n} (x),
\label{eq.59}
\end{equation}
where $N_{n,m}$ is the normalization constant.

In terms of $r$, the radial wave function will be
\begin{equation}
R_{n,m} (r)=N_{n,m}(M\omega\zeta)^{\tau/2}r^{\tau-\frac{1}{2}}e^{-\frac{1}{2}M\omega\zeta r^2}L^{(\tau)}_{n} (M\omega\zeta r^2).
\label{eq.60}
\end{equation}
This normalization constant can be determined by the following condition \cite{EAFB}
\begin{equation}
\frac{1}{b}\,\int\,r^2\,dr\,|R(r)|^2=1.
\label{eq.61}
\end{equation}
Thereby using (\ref{eq.60}), we obtain the normalization constant as follows:
\begin{equation}
N_{n,m}=(2\,M\,\omega\,b\,\zeta)^{1/2}\sqrt{\frac{n!}{(n+\tau)!}}.
\label{eq.62}
\end{equation}

In terms of $x$, one can write the normalized radial wave function
\begin{eqnarray}
&&R_{n,m} (x)=\sqrt{2\,b}\,(M \omega \zeta)^{3/4}\sqrt{\frac{n!}{\Bigg(n+\sqrt{\frac{(|m-\Phi|+\kappa)\,(|m-\Phi|+\kappa+1)}{b^2}+\Big(M\,\omega\,\delta-\frac{1}{2}\Big)^2}\Bigg)!}}\times\nonumber\\
&&x^{\frac{\tau}{2}-\frac{1}{4}}\,e^{-\frac{x}{2}}\,L^{(\tau)}_{n} (x).
\label{eq.63}
\end{eqnarray}

Equation (\ref{eq.58}) is the relativistic energy spectrum and Eq. (\ref{eq.63}) is the normalized radial wave function of the scalar charged bosons via the generalized DKP oscillator equation embedded in the background of a point-like global monopole in the presence of the AB-flux field. One can see that the eigenvalue solutions are influenced by the topological defect parameter $b$ and the Cornell-type potential function Eq. (\ref{eq.19}). This eigenvalue solution is completely different from the result of the DKP equation obtained in \cite{aa1}.

In the function $f(r)$, we choose $\zeta \rightarrow 1$ and $\delta \rightarrow 0$, that is, the DKP oscillator in the point-like global monopole under the AB-flux field. Therefore, the energy eigenvalues from Eq.(\ref{eq.58}) in this special case become
\begin{equation}
\label{eq.64}
E_{n,m}=\pm\,\sqrt{M^2+2\,M\,\omega\,\Bigg(2\,n+\sqrt{\frac{(|m-\Phi|+\kappa)\,(|m-\Phi|+\kappa+1)}{b^2}+\frac{1}{4}}\Bigg)}.
\end{equation}

The normalized radial wave function will be 
\begin{eqnarray}
&&R_{n,m} (x)=\Big[4 b^2 (M \omega)^3\Big]^{1/4}\sqrt{\frac{n!}{\Big(n+\sqrt{\frac{(|m-\Phi|+\kappa)\,(|m-\Phi|+\kappa+1)}{b^2}+\frac{1}{4}}\Big)!}}\times\nonumber\\
&&x^{\frac{\iota}{2}-\frac{1}{4}}\,e^{-\frac{x}{2}}\,L^{(\iota)}_{n} (x),
\label{eq.65}
\end{eqnarray}
where $\iota=\sqrt{\frac{(|m-\Phi|+\kappa)\,(|m-\Phi|+\kappa+1)}{b^2}+\frac{1}{4}}$. 

Equations (\ref{eq.64})--(\ref{eq.65}) is the relativistic eigenvalue solution of the DKP-oscillator in the point-like global monopole space-time background under the influence of the quantum flux field.

\section{Persistent currents of the quantum systems}

In this section, we calculate the persistent currents of the above quantum systems because the eigenvalue solutions obtained in Sect. 3 \& 4 depends on the geometric quantum phase. One can easily show that these energy eigenvalues $E_{n,m}$ is a periodic function of $\Phi_{AB}$ with a periodicity $\Phi_0$, that is, $E_{n,m} (\Phi_{AB}\pm \nu\,\Phi_0)=E_{n,m\mp \nu} (\Phi_{AB})$ with $\nu=0,1,2,3,...$. This dependence of the energy eigenvalue on the geometric quantum phase shows the gravitational analogue of the Aharonov-Bohm effect \cite{YA,MP, ff12, ff13, ff14, ff15, ff16}, quantum mechanical phenomenon and gives rise to a persistent current. It is well-known in condensed matter physics \cite{mm3,mm4} and quantum systems \cite{aa7,mm5,mm6,mm7,nn1} that when there exists dependence of the energy levels on the geometric quantum phase, then, a persistent current arise in the system.

The expression for the total persistent currents is given by \cite{mm5,mm6,mm7}
\begin{equation}
I=\sum_{n,m}\,I_{n,m},
\label{eq.a1}
\end{equation}
where $I_{n,m}$ is called the Byers-Yang (BY) relation defined by
\begin{equation}
I_{n,m}=-\frac{\partial\,E_{n,m}}{\partial\,\Phi_{AB}}.
\label{eq.a2}
\end{equation} 

For the energy expression Eq. (\ref{eq.24}), one will have the following BY-relation
\begin{eqnarray}
I_{n,m}=\mp\,\frac{M\,\omega\,\zeta\,q\,\frac{|m-\Phi|'}{a}\,\Bigg[\frac{|m-\Phi|}{a}+\frac{1}{2}+\kappa\Bigg]} {2\,\pi\,j}\frac{1}{\sqrt{M^2+M\,\omega\,\zeta\,\Bigg(M\,\omega\,\delta+4\,n+2\,j-1\Bigg)}}.
\label{eq.a3}
\end{eqnarray}

Similarly, for the energy expression Eq. (\ref{eq.30}), we have
\begin{eqnarray}
I_{n,m}=\mp\,\frac{M\,\omega\,q\,\frac{|m-\Phi|'}{a}} {2\,\pi\,\sqrt{M^2+2\,M\,\omega\,\Bigg(2n+\frac{|m-\Phi|}{a}+\kappa\Bigg)}}.\quad
\label{eq.a4}
\end{eqnarray}
One can see that persistent currents are influenced by the topological defect characterized by the parameter $a$ of the cosmic string. In addition, the BY-relation given in Eq. (\ref{eq.a3}) is modified by the function $f(r)=\Big(\zeta\,r+\frac{\delta}{r}\Big)$ considered for the generalized DKP oscillator in comparison to that of the expression (\ref{eq.a4}) for the DKP oscillator.

For the energy expression Eq. (\ref{eq.58}) of the generalized DKP oscillator obtained in the background of point-like global monopole, one will have the following BY-relation 
\begin{eqnarray}
I_{n,m}=\mp\,\frac{M\,\omega\,\zeta\,q\,|m-\Phi|'\,\Bigg[|m-\Phi|+\frac{1}{2}+\kappa\Bigg]}{2\,\pi\,b^2\,\tau}\frac{1}{\sqrt{M^2+M\,\omega\,\zeta\,\Bigg(M\,\omega\,\delta+4\,n+2\,\tau-1\Bigg)}}.
\label{eq.a5}
\end{eqnarray}

Similarly, for the energy expression Eq. (\ref{eq.64}), we have
\begin{eqnarray}
&&I_{n,m}=\mp\,\frac{M\,\omega\,q\,|m-\Phi|'\,\Bigg[m-\Phi|+\frac{1}{2}+\kappa\Bigg]} {2\,\pi\,b^2\,\sqrt{M^2+2\,M\,\omega\,\Big(2\,n+\iota\Big)}}.
\label{eq.a6}
\end{eqnarray}
One can see that persistent currents are influenced by the topological defect of a point-like global monopole characterized by the parameter $b$. In addition, the BY-relation given in Eq. (\ref{eq.a5}) is modified by the function $f(r)=\Big(\zeta\,r+\frac{\delta}{r}\Big)$ considered for the generalized DKP oscillator in comparison to that of (\ref{eq.a6}) for the DKP oscillator.

\section{Conclusions}

In this analysis, two specific space-time backgrounds are considered. The first one is a cosmic string space-time described in the spherical system, and the second is a point-like global monopole. In {\tt section 3}, we derived the radial equation of the generalized DKP oscillator equation in cosmic string space-time under the influence of the AB-flux field. After a few mathematical steps, we arrived at a homogeneous second-order differential equation and solved through the parametric Nikiforov-Uvarov method. The relativistic energy eigenvalue is obtained, and in fact it has shown that the topological defects of cosmic string characterized by the parameter $a$ influence the eigenvalue solution and shifted the result compared to the flat space case. There, we discussed a special case corresponds to the DKP oscillator and presented the energy eigenvalue by Eq. (\ref{eq.30}), and the normalized radial wave function by (\ref{eq.31}). We have shown that the eigenvalue solution of the DKP oscillator is modified by the presence of the quantum flux field compared to the result obtained in Ref. \cite{hh6}.

In {\tt section 4}, we derived the radial equation of the same relativistic wave equation in a point-like global monopole under the influence of the AB-flux field. We solved the radial equation through the same technique done earlier, and presented the relativistic energy eigenvalue by Eq. (\ref{eq.58}) and the normalized wave function by Eq. (\ref{eq.63}). We showed that the topological defects of a point-like global monopole characterized by the parameter $b$ influences the eigenvalue solution and shifted the result compared to the flat space case. A special case corresponding to the DKP oscillator was discussed and presented the relativistic energy levels by Eq. (\ref{eq.64}), and the normalized radial wave function by (\ref{eq.65}). 

In both {\tt sections 3-4}, we showed that the energy levels and normalized wave functions are influenced by the AB-flux field and modified the results in addition to the topological defects of the geometries. We have seen that the presented eigenvalue solutions depend on the geometric quantum phase $\Phi_{AB}$, and this dependence of the eigenvalue suggests the existence of the gravitational analogue to the Aharonov-Bohm effect \cite{YA,MP, ff12, ff13, ff14, ff15, ff16}. It is well-known that this dependence of the eigenvalue solution on the quantum phase gives rise to persistent currents that have applications in condensed matter physics as well as other branches of physics. We obtained the Byers-Yang (BY) relation, and showed that the topological defects of cosmic string and point-like global monopole influences this physical quantity. 

In conclusion, our study presented some important results regarding the relativistic quantum systems in the presence of gravitational field effects caused by the topological defects, along with the influence of the quantum flux field. These results hold significant importance and relevance in the current literature. By considering the effects of topological defects on quantum dynamics of the particles, we gained valuable insights into the behavior of quantum systems in curved space-time backgrounds. The interplay between gravity and quantum mechanics is a fundamental aspect of theoretical physics, and connection between these theories are still unclear.

\section*{Acknowledgement}

We sincerely acknowledged the anonymous referee's for their valuable comments and helpful suggestions.

\section*{Conflict of Interest}

Authors declares no conflict of interests.

\section*{Data Availability}

No data generated or analyzed in this paper.

\end{document}